\documentclass[12pt]{article}

\usepackage{amsmath,amssymb,pictex,cite,eepic,epsfig,psfrag,url}
\advance \topmargin by -\headheight
\advance \topmargin by -\headsep
\evensidemargin \oddsidemargin
\makeatletter
\@addtoreset{equation}{section}

\makeatother

\makeatletter
\def\Appendix{\appendix
  \def\@seccntformat##1{Appendix~\csname the##1\endcsname.~~}}
\makeatother

\def\al{
\alpha }

\def\Ga{
\Gamma_{b} }

\def\up{\Upsilon_{b}}

\begin{document}
\hfill \hbox{ITEP-LAT/2009-17}
\vspace{1.5cm}



\bigskip

\begin{center}
{\Large \textbf{Higher Equations of Motion in Boundary Liouville Field Theory}}

\vspace{1.5cm}

{\large A.~Belavin}

\vspace{0.2cm}

Landau Institute for Theoretical Physics, RAS, Chernogolovka, Russia

\vspace{0.2cm}

and

\vspace{0.2cm}

{\large V.~Belavin}

\vspace{0.2cm}
Institute of Theoretical and Experimental Physics and\\ Theory Department, Lebedev Physical Institute, RAS,
Moscow, Russia

\end{center}

\vspace{1.0cm}

\textbf{Abstract}
In addition to the ordinary bulk higher equations of motion in the boundary version of the
Liouville conformal field theory, an infinite set of relations containing the boundary operators
is found. These equations are in one-to-one correspondence with the singular representations
of the Virasoro algebra. We comment on the possible applications in the context of minimal 
boundary Liouville gravity.

\section{Introduction}

Alexey Zamolodchikov showed in~\cite{higher} that a special set of relations
holds for quantum operators in the Liouville conformal field theory (LFT).
They are parameter\-ized by pairs of positive integers $(m,n)$ related
to the degenerate representation of the Virasoro algebra. In the classical limit, they represent
the ``higher equations of mo\-tion'' (HEMs), because the first one $(1,1)$ coincides with the
usual Liouville equation of motion. These equations relate different basic LFT primary fields
$V_{a}(x)$. The equations are derived based on the two main postulates of LFT.
The first is that all singular vectors vanish in the representations built on the exponential
fields; this is a quantum version of the relations in the classical LFT.
The second basically states that the set of exponential fields $\left\{V_{a}\right\}$
(with complex $a$ allowed) in some sense covers the whole variety of primary fields in LFT.

The higher equations turned to be useful in the context of minimal Liouville gravity 
(see~\cite{threep} for the terminology). In particular,
they were used in~\cite{BAl,4pSLG,SupModInt} to derive the general four-point correlation functions
with one degenerate matter field. It is very likely that HEMs are potentially
important in the general program of explicitly constructing the complete set
of correlation functions in the minimal Liouville gravity.

Our purpose here is to reveal an additional set of relations for quantum operators
in LFT with a conformal boundary (BLFT). In the next section, we collect the main
facts about the BLFT. In section 3, we derive the boundary version of HEMs. Possible applications are
considered in section 4. The definitions of the special functions  and also some explicit
calculations omitted in the main text are presented in the appendices.

\section{Boundary Liouville field theory}

We consider the Liouville conformal field theory on a domain $\Gamma$ with a boundary $\partial\Gamma$.
In a general background, the action of the BLFT is~\cite{FZZ}
\begin{equation}
A_{\mathrm{bound}}=\frac{1}{4\pi}\int_{\Gamma}\left[g^{ab}\partial_a\phi\partial_b\phi+QR\phi
+ 4\pi\mu e^{2b\phi}\right]\sqrt{g}d^{2}x+\int\limits_{\partial\Gamma}\left(
\frac{QK}{2\pi}\phi+\mu_{B}e^{b\phi}\right)
\sqrt{g_l} dx \, .\label{gbound} 
\end{equation}
The first term describes the theory in the bulk. Here, $R$ is the scalar curvature associated with the
background metric $g$, and $\mu$ is the bulk cosmological constant. The background charge $Q=b+1/b$
determines the central charge of the theory
\begin{equation}
c_L=1+6Q^2\;.
\end{equation}
A conformally invariant boundary condition~\cite{Cardy} is introduced through the boundary
interaction~\cite{FZZ} given by the second term in~\eqref{gbound}. Here, $g_l$ is the induced 
boundary value of the metric, $K$ is the geodesic curvature of the  boundary, and $\mu_B$ is 
the boundary cosmological constant.
We always imply the upper half-plane geometry in what follows. In the bulk, the holomorphic
component of the stress tensor has the form
\begin{eqnarray}
T(z)&=&-(\partial\phi)^{2}+Q\partial^{2}\phi,
\end{eqnarray}
where $z=x+i y$. Because of the boundary condition on the Liouville field,
the boundary value of the classical stress tensor is
\begin{eqnarray}
T_{\text{cl}}(x)=-\frac 1{16} \varphi_x^2+\frac{1}{4}\varphi_{xx}+\pi b^2(\pi \mu_B^2 b^2-\mu)
e^{\varphi},
\end{eqnarray}
where $\varphi=2 b \phi$. This is equivalent to the classical equation
\begin{eqnarray}
\left(\frac{d^2}{dx^2}+T_{\text{cl}}\right) e^{-\varphi/4}=\pi b^2(\pi \mu_B^2 b^2-\mu)e^{3\varphi/4}
\label{classlim}
\end{eqnarray}
relating two boundary exponents.
This equation is an example of the classical limit of the quantum relations between the
boundary primary operators, which we derive below.

We let $V_{\al}(z,\bar{z})$ denote the bulk primary fields. These fields
have the conformal weight $\Delta_{\alpha}=\alpha(Q-\alpha)$.
The structure constant $C(\al_3,\al_2,\al_1)$ related to the three bulk primaries for
generic values of $\al_i$, $i=1,2,3$, is given~\cite{DO,ZZ} by the expression
\begin{eqnarray}
\lefteqn{C(\al_3,\al_2,\al_1)= \left\lbrack \pi \mu \gamma(b^{2})b^{2-2b^{2}}
\right\rbrack^{\frac{Q-\al_1-\al_2-\al_3}{b}}}. \nonumber \\
&&
\frac{\Upsilon_0\up(2\al_1)\up(2\al_2)\up(2\al_3)}{\up(\al_1+\al_2+\al_3-Q)
\up(\al_1+\al_2-\al_3)\up(\al_1+\al_3-\al_2)\up(\al_2+\al_3-\al_1)},
\label{3pointsZZ}
\end{eqnarray}
where the special function $\up$ and also the other special functions used in the main text
are defined in Appendix A.

The boundary operators $B_{\beta}^{\sigma_{2}\sigma_{1}}(x)$ have the conformal weight
$\Delta_{\beta}=\beta(Q-\beta)$ and are labeled by the two indices
$\sigma_{1}$ and $\sigma_{2}$ related to the left and the right cosmological constants
$\mu_{B_1}$ and $\mu_{B_2}$ by
\begin{eqnarray}
\text{cos}\left(2\pi
b(\sigma-\frac{Q}{2})\right)=\frac{\mu_{B}}{\sqrt{\mu}}\sqrt{\text{sin}(\pi
b^{2})}. \label{mu-sigma}
\end{eqnarray}
The observables depend on the scale-invariant ratios
$\mu_i/\mu_B^2$. For example, the correlation function of $n$ bulk
operators $V_{\alpha_1}\dots V_{\alpha_n}$ and of $m$ boundary operators 
$B_{\beta_1}^{\sigma_{1}\sigma_{2}}\dots B_{\beta_m}^{\sigma_{m}\sigma_{1}}$ 
scales as
\begin{eqnarray}
\mathcal{G}(\alpha_1,\dots\alpha_n,\beta_1\dots\beta_m)\sim
\mu^{(Q-2\sum_i \alpha_i-\sum_j \beta_j)/2b}
F\left(\frac{\mu^2_{B_1}}{\mu},\frac{\mu^2_{B_2}}{\mu},\dots,
\frac{\mu^2_{B_m}}{\mu}\right)\; ,\nonumber
\end{eqnarray}
where $F$ is some scaling function.

To characterize the LFT on the upper half-plane,
we must know~\cite{FZZ} some other structure constants in addition to the bulk
three-point function $C(\al_1,\al_2,\al_3)$:
\begin{enumerate}
\item the bulk one-point function \cite{FZZ}
\begin{equation}
\left\langle V_{\alpha}(z,\bar{z})\right\rangle
=\frac{U(\alpha|\mu_{B})}{\left| z-\bar{z}\right|
^{2\Delta_{\alpha}}}, \label{onepoint} \nonumber
\end{equation}

\item the boundary two-point function \cite{FZZ}
\begin{equation}
\left\langle
B_{\beta_1}^{\sigma_{1}\sigma_{2}}(x)B_{\beta_1}^{\sigma_{2}\sigma_{1}}(0)\right\rangle
=\frac{S(\beta_1,\sigma_{2},\sigma_{1})\delta(\beta_2-\beta_1)}{\left|
x\right|^{2\Delta_{\beta_1}}}, \nonumber
\end{equation}

\item the bulk-boundary two-point function \cite{hosomichi}
\begin{equation}
\left\langle
V_{\alpha}(z,\bar{z})B_{\beta}^{\sigma\sigma}(x)\right\rangle
=\frac{R(\alpha,\beta |\mu_{B})}{\left|  z-\bar{z}\right|
^{2\Delta_{\alpha}-\Delta_{\beta}}\left| z-x\right|
^{2\Delta_{\beta}}}, \label{bbound} \nonumber
\end{equation}
and

\item the boundary three-point function~\cite{PT3}
\begin{eqnarray}
\left\langle
B_{Q-\beta_{3}}^{\sigma_{1}\sigma_{3}}(x_{3})B_{\beta_{2}}^{\sigma_{3}\sigma_{2}}(x_{2})B_{\beta_{1}}^{\sigma_{2}
\sigma_{1}}(x_{1})\right\rangle
=\frac{C_{\beta_{2}\beta_{1}}^{(\sigma_{3}\sigma_{2}\sigma_{1})\beta_{3}}}{\left|
x_{21}\right|  ^{\Delta_{1}+\Delta_{2}-\Delta_{3}}\left|
x_{32}\right|
^{\Delta_{2}+\Delta_{3}-\Delta_{1}}\left|  x_{31}\right|  ^{\Delta_{3}%
+\Delta_{1}-\Delta_{2}}}. \nonumber
\end{eqnarray}
\end{enumerate}
We here give the boundary structure constant explicitly because we use it in the following sections:
\begin{eqnarray}
\lefteqn{C_{\beta_{2}\beta_{1}}^{(\sigma_{3}\sigma_{2}\sigma_{1})\beta_{3}}
=
 \bigl(\pi \mu \gamma(b^2) b^{2-2b^2}\bigr)^{\frac{1}{2b}(\beta_3-\beta_2-\beta_1)}} \nonumber \\
&&
\times\frac{\Gamma_b(2Q-\beta_1-\beta_2-\beta_3)\Gamma_b(\beta_2+\beta_3-\beta_1)
 \Gamma_b(Q+\beta_2-\beta_1-\beta_3)\Gamma_b(Q+\beta_3-\beta_1-\beta_2)}
{\Gamma_b(2\beta_3-Q)\Gamma_b(Q-2\beta_2)\Gamma_b(Q-2\beta_1)\Gamma_b(Q)}
\nonumber \\
&& \quad\times\frac{S_b(\beta_3+\sigma_1-\sigma_3)S_b(Q+\beta_3-\sigma_3-\sigma_1)}{S_b(\beta_2+\sigma_2-\sigma_3)S_b(Q+\beta_2-\sigma_3-\sigma_2)} \nonumber \\
&& \quad \times \frac{1}{i}\int\limits_{-i\infty}^{i\infty}ds \;\;
\frac{S_b(U_1+s)S_b(U_2+s)S_b(U_3+s)S_b(U_4+s)}
{S_b(V_1+s)S_b(V_2+s)S_b(V_3+s)S_b(Q+s)}.  \label{bound3points}
\end{eqnarray}
The coefficients $U_i$ and $V_i$, $i=1,\ldots,4$ are
$$
\begin{array}{ll}
 U_1 =\sigma_1+\sigma_2-\beta_1,            &  V_1 = Q+\sigma_2-\sigma_3-\beta_1+\beta_3, \\
 U_2 = Q-\sigma_1+\sigma_{2}-\beta_1,       &  V_2 = 2Q+\sigma_2-\sigma_3-\beta_1-\beta_3, \\
 U_3 = \beta_2+\sigma_2-\sigma_3,           &  V_3 = 2\sigma_2, \\
 U_4 = Q-\beta_2+\sigma_2-\sigma_3.         \\
\end{array}
$$

\section{Higher equations of motion}

Before discussing the boundary case, we briefly recall the reasoning leading to HEMs in the LFT
without boundary~\cite{higher}. The degenerate primary field $V_{m,n}$
appears for the Kac values~\cite{Kac} of the conformal dimension $\Delta_{m,n}$
related to the parameter
\begin{equation}
\alpha_{m,n}=\frac{Q}2-\frac{\left(  mb^{-1}+nb\right)}2.\label{dmn}%
\end{equation}
We let $D_{m,n}$ denote the singular vector creating operators.
It was proved in~\cite{higher} that the primary field
\begin{equation}
D_{m,n}\bar D_{m,n}V_{m,n}^{\prime}\label{prim}
\end{equation}
can be constructed, where the degenerate logarithmic field $V_{\alpha}^{\prime}$ is defined as
\begin{equation}
V_{m,n}^{\prime}=\frac12\frac{\partial V_{\alpha}}{\partial \alpha}\bigg|_{\alpha=\alpha_{m,n}}.\label{Vprim}%
\end{equation}
Field~\eqref{prim} has the conformal dimension $\Delta_{m,n}+mn$.
It should be identified with the primary field
$V_{m,-n}$ of the same dimension. The operator-valued relation
\begin{equation}
D_{m,n}\bar D_{m,n}V_{m,n}^{\prime}=B_{m,n} V_{m,-n}\label{heq}%
\end{equation}
then holds. The coefficients $B_{m,n}$ are defined explicitly:
\begin{equation}
B_{m,n}=\left(  \pi\mu\gamma(b^{2})\right)  ^{n}b^{1+2n-2m}\gamma
(m-nb^{2})\prod_{\substack{k=1-n \\l=1-m \\(k,l)\neq(0,0) }}^{\substack{m-1
\\n-1 }}(lb^{-1}+kb).\label{klprod}%
\end{equation}

To formulate the boundary analogue of the bulk HEMs, i.e., the relation that should relate the
boundary operators, we first note that the method described above
is no longer applicable. Indeed, for the bulk construction we need both left and right singular 
vector creating operators, while we have only one Virasoro algebra on the boundary. 
Instead we consider the action of $D_{m,n}$ on the primary 
boundary field $B_{m,n}^{s_{1}s_{2}}$ with arbitrary values of the boundary parameters.
The analysis~\cite{FZZ} of classical limit~\eqref{classlim} shows that
this field should not vanish in the general case. On the other hand, it follows for purely algebraic reasons
that the field $D_{m,n} B_{m,n}^{s_{1}s_{2}}$ has the properties of the primary field regardless of the value of
the boundary cosmological parameters. Taking the main assumption of LFT into account, i.e., there exists only one
primary field of a given conformal dimension, we must identify
\begin{align}
D_{m,n} B_{m,n}^{s_{1}s_{2}}=K_{m,n}^{s_{1}s_{2}}B_{m,-n}^{s_{1}s_{2}}
\label{BHEM}
\end{align}
up to a numerical constant, where we have the primary boundary field of the same conformal dimension in the RHS.
This operator-valued relation assumes the corresponding relations between correlation functions if one of the fields
is the subject of~\eqref{BHEM}. In principle, the consistency of this statement must be verified
for arbitrary correlation functions. As usual, it suffices to consider the consequence of~\eqref{BHEM} for the
three-point functions or, equivalently, for the structure constants of the boundary operator product
expansion
\begin{align}
\left\langle B_{\beta_2}^{s_{3}s_{1}}(0)D_{m,n} B_{m,n}^{s_{1}s_{2}}(x)
B_{\beta_1}^{s_{2}s_{3}}(\infty)\right\rangle
=K_{m,n}^{s_{3}s_{2}}
\left\langle B_{\beta_2}^{s_{3}s_{1}}(0)B_{m,-n}^{s_{1}s_{2}}(x)
B_{\beta_1}^{s_{2}s_{3}}(\infty)\right\rangle.
\label{3pHEM}
\end{align}
We use this relation to define the coefficients
\begin{align}
K_{m,n}^{s_{1}s_{2}}=
\frac{\left\langle B_{\beta_2}^{s_{3}s_{1}}(0)D_{m,n} B_{m,n}^{s_{1}s_{2}}(x)
B_{\beta_1}^{s_{2}s_{3}}(\infty)\right\rangle}
{\left\langle B_{\beta_2}^{s_{3}s_{1}}(0)B_{m,-n}^{s_{1}s_{2}}(x)
B_{\beta_1}^{s_{2}s_{3}}(\infty)\right\rangle}.
\label{K}
\end{align}
We note that a rather nontrivial consequence of~\eqref{BHEM} is that the ratio of two correlation
functions~\eqref{K} depends neither on the conformal dimension of the other two fields $\beta_1$ and $\beta_2$
nor on the cosmological parameter $s_3$ of a boundary segment not directly connected  to $B_{m,n}^{s_{1}s_{2}}(x)$.
Before considering the general situation, we test this idea in the case  where the screening
calculations allow avoiding the complicated special functions in expression~\eqref{bound3points}
for the general three-point boundary correlation function.

\subsection{Screening calculations}
In this section, we apply~\eqref{BHEM} in the case where the three-point correlation
functions can be computed perturbatively as Coulomb gas integrals~\cite{FF,DF}.
We recall that if the conformal parameters $\alpha_i$ and $\beta_i$ of the correlation
function $\langle V_{1} \cdots B_1 \cdots\rangle$ satisfy the screening relation
$\sum \alpha_i+\sum \beta_k =Q-n b$, then this correlation
function has a pole~\cite{GL}, and the residue is calculated using the perturbation theory
in $\mu$ and $\mu_B$. Because of the total charge balance condition, only a finite number of terms
in the series have nonzero values. We consider
\begin{align}
D_{1,2} B_{1,2}^{s_{1}s_{2}}=K_{1,2}^{s_{1}s_{2}}B_{1,-2}^{s_{1}s_{2}}.
\end{align}
Taking into account that
$a_{1,2}=-b/2$ and $a_{1,-2}=3 b/2$, we chose the other two fields
such that the screening relation
\begin{align}
\left\langle B_{\beta}^{s_{3}s_{1}}(0)D_{1,2}B_{-b/2}^{s_{1}s_{2}}(x)
B_{Q-\beta-3b/2}^{s_{2}s_{3}}(\infty)\right\rangle
=K_{1,2}^{s_{1}s_{2}}
\left\langle B_{\beta}^{s_{3}s_{1}}(0)B_{3b/2}^{s_{1}s_{2}}(x)
B_{Q-\beta-3b/2}^{s_{2}s_{3}}(\infty)\right\rangle
\end{align}
is satisfied. The action of $D_{1,2}$ reduces to a factor that is known explicitly~\cite{higher}.
The total charge balance is performed for the correlation function in the RHS, and hence
\begin{align}
K_{1,2}^{s_{1}s_{2}}=2(1-2 b\beta)(1-2 b\beta -b^2)
\left\langle B_{\beta}^{s_{3}s_{1}}(0)B_{-b/2}^{s_{1}s_{2}}(1)
B_{Q-\beta-3b/2}^{s_{2}s_{3}}(\infty)\right\rangle.
\label{K120}
\end{align}
A nontrivial check of the general statement should be that
the dependence on $\beta$ and $\mu_3$ in the RHS of~\eqref{K120} cancels.
The volume screening contribution is related to the interaction $-\mu \int d^2 z e^{2 b\phi}$,
while the boundary contribution comes in the second order and requires two boundary screenings $e^{b\phi}$,
\begin{align}
\label{LHScorfun}
&\left\langle B_{\beta}^{s_{3}s_{1}}(0)B_{-b/2}^{s_{1}s_{2}}(1)
B_{Q-\beta-3b/2}^{s_{2}s_{3}}(\infty)\right\rangle=-\mu\int\limits_{\mathrm{Im}z>0}d^{2}z\left\langle
e^{2b\phi(z)}B_{\beta}^{s_{3}s_{1}}(0)B_{-b/2}^{s_{1}s_{2}}(1)B_{Q-\beta
-3b/2}^{s_{2}s_{3}}(\infty)\right\rangle_0\nonumber\\
&+\sum_{i,j}\frac{\mu_{i}\mu_{j}}{2}\int_{C_{i}}%
\int_{C_{j}}dx_{1}dx_{2}\left\langle e^{b\phi(x_{1})}e^{b\phi(x_{2})}B_{\beta
}^{s_{3}s_{1}}(0)B_{-b/2}^{s_{1}s_{2}}(1)B_{Q-\beta-3b/2}^{s_{2}s_{3}}%
(\infty)\right\rangle_0,
\end{align}
where the contours are defined as $C_{1}=(-\infty,0)$, $C_{2}=(0,1)$, and $C_{3}=(1,\infty)$
and $\mu_i$ are the corresponding values of the boundary cosmological constant.
The explicit expressions for the free theory correlation functions are
\begin{align}
\langle e^{2b\phi(z)}B_{\beta}^{s_{3}s_{1}}(0)B_{-b/2}^{s_{1}s_{2}}(1)B_{Q-\beta
-3b/2}^{s_{2}s_{3}}(\infty)\rangle_0 =\left|z\right|^{-4 b\beta}\left|1-z\right|^{2b^{2}}\left|z-\bar{z}\right|^{-2b^{2}}
\end{align}
and
\begin{align}
\langle e^{b\phi(x_{1})}e^{b\phi(x_{2})}B_{\beta}^{s_{3}s_{1}}(0)B_{-b/2}^{s_{1}s_{2}}(1)B_{Q-\beta-3b/2}^{s_{2}s_{3}}%
(\infty)\rangle_0= \nonumber\\ =\left|x_1\right|^{-2 b\beta}\left|x_2\right|^{-2b\beta}\left|1-x_{1}\right|^{b^2}
\left|1-x_{2}\right|^{b^{2}}\left|  x_{1}-x_{2}\right|^{-2b^{2}}.
\end{align}
We introduce the notation
\begin{eqnarray}\nonumber
&I(A,B,C)=\int\limits_{\mathrm{Im}z>0}d^{2}z\left|z\bar{z}\right|^{A}\left|(1-z)(1-\bar{z})\right|
^{B}\left|z-\bar{z}\right|  ^{C},\\
&I_{ij}(A,B,C)=\int_{C_{i}}\int_{C_{j}}dx_{1} dx_{2}
 \left|x_1\right|^{A}\left|x_2\right|^{A}\left|1-x_{1}\right|^{B}
\left|1-x_{2}\right|^{B}\left|  x_{1}-x_{2}\right|^{C}.
\end{eqnarray}
We can  write
\begin{align}
\langle B_{\beta}^{s_{3}s_{1}}(0) & B_{-b/2}^{s_{1}s_{2}}(x)
B_{Q-\beta-3b/2}^{s_{2}s_{3}}(\infty)\rangle =\nonumber\\
&=-\mu I(-2b\beta,b^2,-2b^2)+
\sum_{i,j}\frac{\mu_{i}\mu_{j}}2 I_{ij}(-2b\beta,b^2,-2b^2).
\end{align}
All integrations can be performed explicitly. Using the results in Appendices C and D,
we obtain
\begin{align}\nonumber
&I=-\frac{1}{2\pi^3}\sin(\frac{\pi C}{2})\sin(\pi A)\sin(\pi B)\sin(\pi (A+B+C)) J(A,B,C),\nonumber\\
&I_{11}=-\frac{1}{\pi^3} \sin\frac{\pi C}{2} \sin\pi(A+B+\frac C2)\sin\pi(A+B+C) J(A,B,C), \nonumber\\
&I_{12}=-\frac{1}{2\pi^3} \sin\pi C \sin\pi(A+B+\frac C2)\sin\pi(A+\frac C2) J(A,B,C), \nonumber\\
&I_{13}=-\frac{1}{2\pi^3} \sin\pi C \sin\pi(A+B+\frac C2)\sin\pi(B+\frac C2) J(A,B,C),  \\
&I_{22}=-\frac{1}{\pi^3} \sin\frac{\pi C}{2} \sin\pi(A+\frac C2)\sin\pi A J(A,B,C), \nonumber\\
&I_{23}=-\frac{1}{2\pi^3} \sin\pi C \sin\pi(A+\frac C2)\sin\pi(B+\frac C2) J(A,B,C), \nonumber\\
&I_{33}=-\frac{1}{\pi^3} \sin\frac{\pi C}{2} \sin\pi(B+\frac C2)\sin\pi B J(A,B,C),\nonumber
\end{align}
where
\begin{eqnarray}
&J(A,B,C)=\Gamma(A+1)\Gamma(B+1)\Gamma(C+1)\Gamma(-C/2)\Gamma(B+C/2+1)\times\nonumber\\
&\Gamma(-A-B-C-1)\Gamma(-A-B-C/2-1)\Gamma(A+C/2+1).
\end{eqnarray}
Summing these contributions, we obtain
\begin{align}
\langle B_{\beta}^{s_{3}s_{1}}(0) B_{-b/2}^{s_{1}s_{2}}(x)&
B_{Q-\beta-3b/2}^{s_{2}s_{3}}(\infty)\rangle=\nonumber\\
-\frac{1}{2\pi^3}\sin\frac{\pi C}{2}\bigg[&-\mu\sin(\pi A)\sin(\pi B)\sin(\pi (A+B+C))+\nonumber\\
&+ \mu_1^2 \sin(\pi (A+B+C/2))\sin(\pi (A+B+C))+\nonumber\\
&+ \mu_2^2 \sin(\pi (A+C/2))\sin(\pi A)+ \mu_3^2 \sin(\pi (B+C/2))\sin(\pi B)-\\
&-2\mu_1\mu_2\cos\frac{\pi C}{2} \sin(\pi (A+B+C/2))\sin(\pi (A+C/2))-\nonumber\\
&-2\mu_1\mu_3\cos\frac{\pi C}{2} \sin(\pi (A+B+C/2))\sin(\pi (B+C/2))+\nonumber\\
&+2\mu_2\mu_3\cos\frac{\pi C}{2} \sin(\pi (A+C/2))\sin(\pi (B+C/2)) \bigg] J(A,B,C), \nonumber
\end{align}
where $A=-2 b\beta$, $B=b^2$, and $C=-2b^2$. It is sufficiently remarkable that because of the relation
$C=-2B$, the result is independent of $\mu_3$,
\begin{align}
\langle B_{\beta}^{s_{3}s_{1}}(0)B_{-b/2}^{s_{1}s_{2}}(x)
B_{Q-\beta-3b/2}^{s_{2}s_{3}}(\infty)\rangle=
&\frac{1}{2\pi^3}\bigg[-\mu\sin(\pi B)+\mu_1^2+\mu_2^2-2\mu_1\mu_2\cos(\pi B)\bigg]\nonumber\\
&\times\sin(\pi A)\sin(\pi B) \sin(\pi(A-B)) J(A,B,C).
\end{align}
With the FZZ parameterization
\begin{align}
\mu_i^2=\mu \frac{\cosh^2\pi b s_i}{\sin\pi b^2}\,,\,\,\,\,
\mu_i\mu_j=\mu \frac{\cosh\pi b s_i \cosh\pi b s_i}{\sin\pi b^2}\,,
\label{FZZparam}
\end{align}
correlation function~\eqref{LHScorfun} is
\begin{align}
\langle B_{\beta}^{s_{3}s_{1}}(0)&B_{-b/2}^{s_{1}s_{2}}(x)
B_{Q-\beta-3b/2}^{s_{2}s_{3}}(\infty)\rangle=\nonumber\\
&\frac{1}{2\pi^3}
\sin\frac{B+i s_1-i s_2}2\sin\frac{B-i s_1+i s_2}2\sin\frac{B+i s_1+i s_2}2
\sin\frac{B-i s_1-i s_2}2
\nonumber\\
&\times\sin(\pi A)\sin(\pi B) \sin(\pi(A-B)) J(A,B,C).
\end{align}
It can be easily verified that $\beta$-dependence of $K_{1,2}^{s_{1}s_{2}}$ also vanishes. 
Finally,
\begin{align}\label{k12}
&K_{1,2}^{s_{1}s_{2}}= \frac{4\mu \gamma(b^2)}{\pi}\Gamma(1-2b^2)
\Gamma(1-b^2)\Gamma(1+b^2)\times\nonumber\\
&\sin\pi b\frac{b+i(s_1+s_2)}2\sin\pi b\frac{b-i(s_1+s_2)}2\sin\pi b\frac{b+i( s_2-s_1)}2
\sin\pi b\frac{b-i(s_2-s_1)}2.
\end{align}
A similar calculation in the case $(1,1)$ gives
\begin{align}\label{k11}
K_{1,1}^{s_{1}s_{2}}=\left(\frac{4\mu\gamma(b^2)}{\pi}\right)^{1/2}\Gamma(1-b^2)
\sin\pi b\frac{i(s_1+s_2)}2\sin\pi b\frac{i(s_2-s_1)}2.
\end{align}
In the next section, we show that~\eqref{k12} and~\eqref{k11}
are generalized for the general case $(m,n)$.

\subsection{General three-point analysis}
Operator-valued relation~\eqref{BHEM} means that the equality~\eqref{3pHEM}
holds for the general three-point correlation functions. In terms of the boundary structure 
constants~\eqref{bound3points} this gives the following expression for the coefficients
\begin{align}
K_{m,n}^{s_{3}s_{2}}=P_{m,n}(Q-\beta_3-\beta_1)P_{m,n}(\beta_3-\beta_1)
\frac{C_{\beta_{m,n},\beta_1}^{(s_3s_2s_1)\beta_3}}
{C_{\beta_{m,n+n b},\beta_1}^{(s_3s_2s_1)\beta_3}},
\label{kappagener}
\end{align}
where $P_{m,n}$ is the fusion polynomial
\begin{equation}
P_{m,n}(x)=\prod_{\substack{k=1-n :2:n-1 \\l=1-m :2: m-1  }}
(x-\lambda_{l,k}),\label{fusiopol}%
\end{equation}
and $\lambda_{l,k}=(lb^{-1}+kb)/2$.
To calculate the ratio, we use the following generalizations of the shift relations for the
$S_b$ and $\Gamma_b$ functions presented in Appendix A:
\begin{align}
S_b(x+ n b)=2^n\prod_{k=0}^{n-1} \sin \pi b (x+k n) \cdot S_b(x)
\end{align}
and
\begin{align}
\Gamma_b(x+ n b)=
\frac{(2\pi)^{\frac n2} b^{n(b x-\frac 12)}b^{\frac{n(n-1))}2b^2}}
{\prod_{k=0}^{n-1}\Gamma\left[b(x+k b)\right]} \Gamma_b(x).
\label{gammaKN}
\end{align}
It is convenient to split the ratio into three parts. The first contains the ratio of the
integral parts in expression~\eqref{bound3points}. Here, the diference comes from two
$S_b$ functions in the integral. Keeping in mind that $S_b(Q-x)=1/S_b(x)$, we obtain
\begin{align}
S_b(\beta_{mn}+n b +\sigma_2-\sigma_3+s)S_b(Q-\beta_{mn}-n b+\sigma_2-\sigma_3+s)=
\nonumber\\
\prod_{k=0}^{n-1}
\frac{\sin \left[ \pi b \left(\frac{(1+2 k-n)b}{2}+\sigma_2-\sigma_3+s\right)
+\frac{1-m}2 \pi\right]} {\sin \left[-\pi b \left(\frac{(1+2 k-n)b}{2}+\sigma_2-\sigma_3+s\right)
+\frac{1-m}2 \pi\right]}\nonumber\\
\cdot S_b(\beta_{mn} +\sigma_2-\sigma_3+s)S_b(Q-\beta_{mn}+\sigma_2-\sigma_3+s)\nonumber\\
=(-1)^{mn} S_b(\beta_{mn} +\sigma_2-\sigma_3+s)S_b(Q-\beta_{mn}+&\sigma_2-\sigma_3+s).
\end{align}
Hence, the ratio of the integrals gives just $(-1)^{m n}$. The second part of the ratio is the part
containing the $S_b$ functions in the prefactor in front of integral~\eqref{bound3points}.
It contributes
\begin{align}
\frac{S_b(\beta_{mn}+n b+\sigma_2-\sigma_3)S_b(Q+\beta_{mn}+n b -\sigma_2-\sigma_3)}
{S_b(\beta_{mn}+\sigma_2-\sigma_3)S_b(Q+\beta_{mn}-\sigma_2-\sigma_3)}=
\nonumber\\
2^{2n} \prod_{k=0}^{n-1}
\sin \left[ \pi b \left(\frac{(1+2 k-n)b}{2}+\sigma_2-\sigma_3\right)
+\frac{1-m}2 \pi\right]\nonumber \\ \cdot
 \sin \left[\pi b \left(\frac{(3+2 k-n)b}{2}-\sigma_2-\sigma_3\right)
+\frac{3-m}2 \pi\right]
\end{align}
to the ratio. In the FZZ parameterization~\eqref{FZZparam} parameter $\sigma=Q/2+i s/2$ 
and we have
\begin{align}
2^{2n} \prod_{k=0}^{n-1}
\sin \left[ \pi b \frac{(1-m)b^{-1}+(1+2 k-n)b+i(s_2-s_3)}{2}\right]
\nonumber \\ \cdot
\sin \left[ \pi b \frac{(1-m)b^{-1}+(1+2 k-n)b+i(s_2+s_3)}{2}\right].
\end{align}
The third part of the ratio, the part containing the $\Gamma_b$ functions in the prefactor, contributes
\begin{align}
M=\frac
{\Gamma_b(2Q-\beta_1-\beta_{mn}-\beta_3)
\Gamma_b(\beta_{mn}+\beta_3-\beta_1)}
{\Gamma_b(2Q-\beta_1-\beta_{mn}-n b-\beta_3)
\Gamma_b(\beta_{mn}+n b+\beta_3-\beta_1)}
\nonumber \\ \cdot
\frac{\Gamma_b(Q+\beta_{mn}-\beta_1-\beta_3)\Gamma_b(Q+\beta_3-\beta_{mn}-\beta_1)\Gamma_b(Q-2\beta_{mn}-2 n b)}
{\Gamma_b(Q+\beta_{mn}+n b-\beta_1-\beta_3)\Gamma_b(Q+\beta_3-\beta_{mn}-n b-\beta_1)\Gamma_b(Q-2\beta_{mn})}.
\end{align}
The dependence of $M$ on $\beta_1$ and $\beta_3$ comes from two factors of the form
\begin{align}
\frac
{\Gamma_b(\beta_{mn}+\beta)
\Gamma_b(Q-\beta_{mn}+\beta)}
{\Gamma_b(\beta_{mn}+n b-\beta)
\Gamma_b(Q-\beta_{mn}-n b+\beta)}=\frac{b^{-n b(2\beta_{mn}+n b -Q)} b^{-mn}}
{p_{mn}(\beta)}
\end{align}
with $\beta$ equal to either $\beta_3-\beta_1$ or $Q-\beta_3-\beta_1$, and hence
\begin{align}
M=b^{-2nb(2\beta_{mn}+n b -Q)}b^{-2mn}\frac{1}{P_{mn}(\lambda_3-\lambda_1)P_{mn}(\lambda_3+\lambda_1)}
\frac{\Gamma_b(Q-2\beta_{mn}-2 n b)}{\Gamma_b(Q-2\beta_{mn})}.
\end{align}
Combining all together, we obtain
\begin{align}\label{kmn}
&K_{m,n}^{s_{1}s_{2}}= (-1)^{m n}\left(\frac{4 \mu \gamma(b^2)}\pi\right)^{\frac{n}{2}}
 b^{2n(1-m)}
\prod_{k=0}^{2n-1} \Gamma(m-(n-k)b^2)\times\\
&\prod_{k=0}^{n-1}
\sin\pi b\frac{(1-m)b^{-1}+(1+2k-n)b+i(s_1+s_2)}2
\sin\pi b\frac{(1-m)b^{-1}+(1+2k-n)b+i(s_2-s_1)}2.\nonumber
\end{align}
It is easy to see that the degenerate field $B_{m,n}^{s_1,s_2}$ has a vanishing singular vector
and the truncated operator product expansion if $s_1\pm s_2=2 i \lambda_{k,r}$ with
$k=1-n,3-n,\ldots,n-1$ and $r=1-m,3-m,\ldots,m-1$. This result generalize $(1,2)$ fusion rules 
suggested in~\cite{FZZ}.

\section{Application in minimal gravity}

One possible application of the BHEM in a physical context is for constructing the correlation
functions of physical fields in boundary minimal Liouville gravity (BMLG).
This is an alternative description of non-critical open string~\cite{Polyakov} propagating in
``low-dimensional'' space-time. In BMLG,
the gravity is induced by one of the minimal CFT models, and it is expected to be exactly
solvable. This is confirmed by the fact that the alternative approach of matrix models to 2D gravity
(see, e.g.,~\cite{Ginsparg} for a review) provides explicit expressions for many observables. 
The comparison of some open string amplitudes, derived both in matrix model framework and using
worldsheet description, was performed recently in ref.~\cite{hosomichi2}.
The BMLG
theory consists of the matter, Liouville, and ghost sectors, which do not interact except
through the conformal anomaly and through the constriction of the physical fields.
The matter central charge is defined by the central charge balance condition
\begin{equation}
c_{\text{M}}+c_{\text{L}}=26.\label{cadd}%
\end{equation}
The physical fields are defined in the framework of BRST quantization as cohomologies with respect to
the BRST charge
\begin{equation}
\mathcal{Q}=\oint\left(  C(T_{\text{L}}+T_{\text{M}})+C\partial CB\right)
\frac{dz}{2\pi i}.\label{brst}%
\end{equation}
Here, $B$ and $C$ are ghost fields of the respective spins 2 and -1(we use the uppercase letters
for the ghosts here in order not to confuse $B$ with the Liouville parameter $b$).

It is the specific property of MLG
that in the construction of the physical fields of nonzero ghost number all matter fields are ``dressed'' by
Liouville exponentials in the LHS of the HEMs.
For example, there exist ghost number-1 basic boundary physical fields of the form
\begin{equation}
W^{(\alpha_1,\alpha_2|s_1,s_2)}_{m,n} = U_{m,n}^{(\alpha_1,\alpha_2|s_1,s_2)} C
\end{equation}
and
\begin{equation}
\Psi^{\alpha_1,\alpha_2}_{m,n} B^{s_1,s_2}_{m,-n}.
\end{equation}
Here, the parameters $\alpha_1,\alpha_2$ and $s_1,s_2$ correspond to the conformal boundary conditions
to the left/to the right from the operator insertion in the respective matter and Liouville sectors.
Because of the anomaly of the ghost current, the MLG correlation function of any number $N$ of fields
must be of the form~\cite{pol}
\begin{equation}
G_{N}=\prod_{i=4}^N \int d x_i \langle W_1(x_1)W_2(x_2)W_3(x_3) U_4 (x_4) \dots U_N(x_N)\rangle_{\text{MG}},
\label{intMLG}
\end{equation}
where $\left\langle \ldots\right\rangle _{\text{MG}}$ denotes the joint
correlation function of matter, Liouville, and ghosts.

Another important class of boundary physical fields (ghost number $0$) is the boundary ground
ring. The general form of the elements of the boundary ground
ring~\cite{GroundRing1,GroundRing2} (also see~\cite{GroundRing3} for more recent developments) is
\begin{equation}
O_{m,n}=H_{m,n}\Psi_{m,n}V_{m,n}.\label{Omn}%
\end{equation}
Here, $H_{m,n}$ are operators of level $mn$ and ghost number $0$ constructed from the Virasoro
generators
$L_{n}^{\text{(L)}}$, $L_{n}^{\text{(M)}}$, and ghosts. It can be shown analogously to~\cite{BBSLG}
that the boundary higher equations lead to the following important relation
between the two types of physical fields introduced above:
\begin{equation}
\mathcal{Q} O_{m,n}^{s_1,s_2}=  K_{m,n}^{s_1,s_2} W_{mn}^{s_1,s_2}.
\end{equation}
In particular, this means that to construct the boundary ground ring element $O_{m,n}^{s_1,s_2}$,
the fusion relations for the cosmological constants $s_1$ and $s_2$ in the Liouville sector
should be satisfied, $K_{m,n}^{s_1,s_2}=0$. Another consequence of this relation is that if the fusion
rules for $s_1$ and $s_2$ are not satisfied, then the field $W_{m,n}^{s_1,s_2}$ seems exact,
and the correlation functions of this field are naively equal to zero.
This statement should be checked more carefully.

Taking the commutation relations $\{B_{-1},\mathcal{Q}\}=\partial$ into account, we can
straightforwardly verify that
\begin{equation}
U_{m,n}^{(\alpha_1,\alpha_2|s_1,s_2)}=\frac{1}{K_{m,n}^{s_1, s_2}}
\left( \partial-\mathcal{Q} B_{-1}\right)O_{m,n}^{(\alpha_1,\alpha_2|s_1,s_2)} 
\end{equation}
This relation allows performing every
(integrated) insertion of the particular operator $U_k(x_k)$ explicitly,
integrating by parts. The integral~\eqref{intMLG} can thus be reduced to boundary terms, which are in principle defined
by the operator product expansions of the ground ring elements and the basic boundary fields $W_a$.

\section*{Acknowledgments}
The authors thank V.~Fateev, S.~Rebault, I.~Kostov and K~.Hosomichi for the useful discussions. V.~B.~acknowledges
the hospitality of the LPTA of University Montpellier II. This research was conducted in part within the
framework of the federal program ``Scientific and Scientific-Pedagogical Personnel of Innovational  Russia'' 2009-2013
(State Contract No.~P1339) and also by the RFBR initiative interdisciplinary project (Grant No.~09-02-12446-ofi-m)
and by an RBRF-CNRS project (Grant No.~09-02-93106).

\Appendix

\section{Special functions}

Here we collect the definitions and some properties of the special functions $\Gamma_b$, $\up$ 
and $S_b$. 
The Double Gamma function introduced by Barnes \cite{Barnes} is
defined as
\begin{eqnarray}
\nonumber \\
&&
\text{log}\Gamma_{2}(s|\omega_1,\omega_2)=\left(\frac{\partial}{\partial
t}
\sum_{n_1,n_2=0}^{\infty}(s+n_1\omega_1+n_2\omega_2)^{-t}\right)_{t=0}.
\nonumber
\end{eqnarray}
The Barnes Gamma function is defined as
\begin{eqnarray}
\Gamma_b(x) \equiv \frac{\Gamma_2(x|b,b^{-1})}{\Gamma_2(Q/2|b,b^{-1})}.
\end{eqnarray}
The function $\Ga(x)$  satisfies the functional relations
\begin{eqnarray}
\nonumber \\
&&\Ga(x+b)= \frac{\sqrt{2\pi}b^{bx-\frac{1}{2}}}{\Gamma(bx)}\Ga(x), \nonumber \\
&&\Ga(x+1/b)=
\frac{\sqrt{2\pi}b^{-\frac{x}{b}+\frac{1}{2}}}{\Gamma(x/b)}\Ga(x)
\nonumber
\end{eqnarray}
and is a meromorphic function of $x$, whose poles are located at
$x=-nb-mb^{-1}, n,m \in \mathbb{N}$.
The integral representation converges for $0<\mathrm{Re}x$:
\begin{eqnarray}
&&\text{log}\Ga(x)=\int_{0}^{\infty}\frac{dt}{t}\left\lbrack\frac{e^{-xt}-e^{-Qt/2}}{(1-e^{-bt})(1-e^{-t/b})}
-\frac{(Q/2-x)^{2}}{2}e^{-t}-\frac{Q/2-x}{t}\right\rbrack.\nonumber
\end{eqnarray}
The Barnes Sine function is defined as
\begin{eqnarray}
S_b(x)\equiv \frac{\Ga(x)}{\Ga(Q-x)}.
\end{eqnarray}
The shift relations are
\begin{eqnarray}
&& S_b(x+b) = 2\text{sin}(\pi bx)S_b(x),  \nonumber \\
&& S_b(x+1/b) = 2\text{sin}(\pi x/b)S_b(x), \nonumber
\end{eqnarray}
and $S_b(x)$ is a meromorphic function of $x$, whose poles are located
at $x=-nb-mb^{-1}, n,m \in \mathbb{N}$,
and whose zeros are located at $x=Q+nb+mb^{-1}, n,m \in \mathbb{N}$.
The integral representation converges in the strip $0<\mathrm{Re}x<Q$:
\begin{eqnarray}
&&\text{log}S_b(x)=\int_{0}^{\infty}\frac{dt}{t}\left\lbrack\frac{\text{sinh}(\frac{Q}{2}-x)t}
{2\text{sinh}(\frac{bt}{2})\text{sinh}(\frac{t}{2b})}-
\frac{(Q-2x)}{t}\right\rbrack.\nonumber
\end{eqnarray}
Finally, the upsilon function is defined as
\begin{eqnarray}
\up(x)^{-1} \equiv \Ga(x)\Ga(Q-x).
\end{eqnarray}
The functional relations are
\begin{eqnarray}
&&\up(x+b)=\frac{\Gamma(bx)}{\Gamma(1-bx)}b^{1-2bx}\up(x),\nonumber \\
&&\up(x+1/b)=\frac{\Gamma(x/b)}{\Gamma(1-x/b)}b^{2x/b-1}\up(x),
\nonumber
\end{eqnarray}
and $\up(x)$ is an entire function of $x$ whose zeros are located at $x=-nb-mb^{-1}$ and $x=Q+nb+mb^{-1}$,
 $n,m \in \mathbb{N}$.
The integral representation converges in the strip $0<\mathrm{Re}x<Q$:
\begin{eqnarray}
&&\text{log}\up(x)=\int_{0}^{\infty}\frac{dt}{t}\left\lbrack\left(\frac{Q}{2}-x\right)^{2}e^{-t}-
\frac{\text{sinh}^{2}(\frac{Q}{2}-x)\frac{t}{2}}{\text{sinh}\frac{bt}{2}\text{sinh}\frac{t}{2b}}
\right\rbrack. \nonumber
\end{eqnarray}

\section{Contour calculus. Bulk term}
We consider the integral
\begin{align}
I=\int_{\text{Im} z>0} d^2 z (z\bar z)^A\left[(1-z)(1-\bar z)\right]^B |z-\bar z|^C.
\end{align}
It is convenient to extend the integration area to be the whole complex plane:
\begin{align}
I=\frac{2^C}{1+e^{i\pi C}}
\int_{-\infty}^{\infty} d x d y (x^2+y^2)^A\left[(1-x)^2+y^2\right]^B y^C.
\end{align}
For practical computations, it is convenient to decompose this integral into a sum
of holomorphic and antiholomorphic parts using the way proposed in~\cite{Dots1,Dots2}. Performing the Wick rotation $y\rightarrow i y(1-2 i \epsilon)$
(with infinitely small $\epsilon>0$) and introducing the new variables $u=x-y$ and $v=x+y$, we can write
the integral in the factored form
\begin{align}
I=\frac{i^{C+1}}{2(1+e^{i\pi C})}
\int_{-\infty}^{\infty} d v \int_{-\infty}^{\infty}  d u
(v-i (v-u)\epsilon)^A(u+ i (v-u)\epsilon)^A \nonumber\\ \cdot(1-u- i (v-u)\epsilon)^B
(1-v+ i (v-u)\epsilon)^B (v-u- i (v-u)\epsilon)^C.
\end{align}
There are three branch points $u=-i \epsilon v$, $u=1-i(v-1)\epsilon$, and $u=v+i\epsilon$ in the $u$ plane.
By deforming the integration contour, we can easily demonstrate that nontrivial contributions
come only from two domains of integration, namely,
when  $0<v<1$ and $v>1$,
\begin{align}
I=\frac{i^{C+1}}{2(1+e^{i\pi C})}\left[I_1+I_2\right],
\end{align}
where
\begin{align}
I_1&=\int_0^1 d v v^A(1-v)^B\bigg \{e^{i\pi C}\int_v^1 d u u^A(1-u)^B(u-v)^C
+e^{-i\pi C} \int_1^{\infty} d u u^A(1-u)^B(u-v)^C
\nonumber\\
 &+ e^{i\pi (C+B)} \int_1^{\infty} d u u^A(u-1)^B(u-v)^C
+e^{-i\pi (C+B)} \int_{\infty}^1 d u u^A(u-1)^B(u-v)^C \bigg\},
\nonumber\\
I_2&=e^{i\pi B}\int_1^{\infty} d v v^A(1-v)^B
\bigg \{e^{i\pi(-B+ C)}\int_v^{\infty} d u u^A(u-1)^B(u-v)^C\nonumber\\
&+e^{i\pi(-B-C)} \int_{\infty}^v d u u^A(u-1)^B(u-v)^C \bigg\}.
\end{align}
In a more compact form,
\begin{align}
I_1&=2 i \sin \pi C \int_0^1 d v v^{2A+C+1}(1-v)^B \int_0^1 d u u^A(1-u)^C(1-v u)^B+
\nonumber\\
&+2 i \sin \pi (C+B) \int_0^1 d v v^A(1-v)^B \int_0^1 d u u^{-A-B-C-2}(1-u)^B(1-v u)^C,
\nonumber\\
I_2&=2 i \sin \pi C \int_0^1 d v v^{-2A-2B-C-3}(1-v)^B \int_0^1 d u u^{-A-B-C-2}(1-u)^C(1-v u)^B.
\end{align}
We obtain the expression
\begin{align}
I=-\frac{1}{2} \left(\cos\frac{\pi C}{2}\right)^{-1}
\bigg[&\sin\pi(C+B)I_0(A,B,-A-B-C-2,B,C)+\nonumber\\
&\sin\pi C I_0(2A+C+1,B,A,C,B)+\nonumber\\
&\sin\pi C I_0(-2A-2B-C-3,B,-A-B-C-2,C,B)
\bigg],
\label{I0}
\end{align}
where
\begin{align}
I_0(\alpha,\beta,\gamma,\delta,\epsilon)=\int_0^1 d v v^{\alpha} (1-v)^{\beta}
\int_0^1 d u u^{\gamma}(1-u)^{\delta}(1-v u)^{\epsilon}.
\end{align}
We now use standard formulas for the integral representations of the higher hypergeometric functions
\begin{align}
I_0(\alpha,\beta,\gamma,\delta,\epsilon)=B(\gamma+1,\delta+1)
\int_0^1 d v v^{\alpha} (1-v)^{\beta}
{}_2 F_{1}(-\epsilon,\gamma+1,\gamma+\delta+2;x)=\nonumber\\
B(\gamma+1,\delta+1)B(\alpha+1,\beta+1)
{}_3 F_{2}(-\epsilon,\gamma+1,\alpha+1;\gamma+\delta+2,\alpha+\beta+2;1),
\label{J}
\end{align}
where $B(x,y)$ is the beta function. In our cases, the integral $I_0$ appears at special values of the
parameters and can be expressed in terms of gamma functions because of the formula
\begin{align}
{}_3 F_{2}(a,b,c;a-b+1,a-c+1;1)=
\frac{\Gamma(1+a/2)\Gamma(1+a-b)\Gamma(1+a-c)\Gamma(1+a/2-b-c)}
{\Gamma(1+a)\Gamma(1+a/2-b)\Gamma(1+a/2-c)\Gamma(1+a-b-c)}.
\end{align}
Thus,
\begin{align}\label{JJJ}
&I_0(A,B,-A-B-C-2,B,C)=\frac1{2\pi^3} \sin \pi C \sin \pi(A+B+\frac C2)\sin\pi(A+\frac C2) \cdot J(A,B,C),\nonumber\\
&I_0(2A+C+1,B,A,C,B)=-\frac1{2\pi^3} \sin \frac{\pi C}2 \sin \pi(A+B+\frac C2)\sin\pi (A+B+C) \cdot J(A,B,C), \\
&I_0(-2A-2B-C-3,B,-A-B-C-2,C,B)=-\frac1{2\pi^3} \sin \frac{\pi C}2 \sin \pi(A+\frac C2)\sin\pi A \cdot J(A,B,C),\nonumber
\end{align}
where
\begin{align}
J(A,B,C)=\Gamma(A+1)\Gamma(B+1)\Gamma(C+1)\Gamma(-\frac C2)\Gamma(-A-B-C-1)
\Gamma(-A-B-\frac C2-1)\nonumber\\
\times\Gamma(B+\frac C2+1)\Gamma(A+\frac C2+1)
\end{align}
Substituting~\eqref{JJJ} in~\eqref{I0} gives
\begin{align}
I=-\frac{1}{4\pi^3} \left(\cos\frac{\pi C}{2}\right)^{-1}
\sin\pi A\sin\pi B\sin\pi C\sin\pi(A+B+C) J(A,B,C).
\end{align}

\section{Contour calculus. Boundary terms}
Here, we give the details of the calculations of the integrals that appear when calculating
the coefficient $K_{12}$. The parameterization of the boundary is chosen as follows.
The intervals $(-\infty,0)$, $(0,1)$, and $(1,\infty)$ are respectively denoted by $C_1$, $C_2$,
and $C_3$. The corresponding cosmological parameters are $s_3$, $s_1$, and $s_2$.
We are interested in the integrals
\begin{align}
I_{ij}=\int_{C_i} dx_1 \int_{C_j} dx_2 |x_1|^A|x_2|^A|1-x_1|^B|1-x_2|^B|x_1-x_2|^C.
\end{align}
With $I_{ij}=I_{ji}$ taken into account, it suffices to consider $i\leq j$.
Below, we often use the result for the two-dimensional Selberg integral
\begin{align}
\int_0^1 dx_1 \int_{0}^1 dx_2 |x_1|^{\mu-1}|x_2|^{\mu-1}|1-x_1|^{\nu-1}|1-x_2|^{\nu-1}
|x_1-x_2|^{2 g}=\nonumber\\
=2\frac{\Gamma(\mu)\Gamma(\nu)\Gamma(\mu+g)\Gamma(\nu+g)}
{\Gamma(g)\Gamma(\mu+\nu+g)\Gamma(\mu+\nu+2g)}.
\end{align}
The first integral has the form of the Selberg integral and can be written as
\begin{align}
I_{11}=-\frac{1}{\pi^3} \sin\frac{\pi C}{2} \sin\pi(A+B+\frac C2)\sin\pi(A+B+C) N.
\end{align}
The integral $I_{12}$ reduces to the function $I_0$ introduced in~\eqref{J} (in the previous appendix)
after $x_2\rightarrow1/x_2$,
\begin{align}
I_{12}=I_0(A,B,-A-B-C-2,B,C).
\end{align}
To calculate $I_{13}$, we consider the contour integral in the complex plane $x_2$ such that
the contour goes along the real axis above the branch points $0$, $x_1$, and $1$.
The contour can be moved to infinity, which gives the relation
\begin{align}
I_{13}+e^{-i\pi A} \int_0^1 \int_0^{x_1}+e^{-i\pi(A+C)} \int_0^1 \int_{x_1}^1+
e^{-i\pi(A+C+B)} \int_0^1 \int_1^{\infty}=0
\end{align}
or
\begin{align}
I_{13}+e^{-i\pi A} \int_0^1 \int_0^{x_1}+e^{-i\pi(A+C)}
\bigg[I_{11}-\int_0^1 \int_0^{x_1}\bigg]+
e^{-i\pi(A+C+B)} I_{12}=0.
\end{align}
In the interval $(0,x_1)$, the integral coincides with $I_0(2A+C+1,B,A,C,B)$. Taking into account that
all integrals are real, we obtain
\begin{align}
I_{13}=-\cos\pi(A+C) I_{11}-\cos\pi(A+B+C) I_{12}+\nonumber\\ +[\cos\pi(A+C)-\cos\pi A]
I_0(2A+C+1,B,A,C,B).
\end{align}
Integral $I_{22}$ reduces to the Selberg integral after $x_2\rightarrow 1/x_2$
and $x_2\rightarrow 1/x_2$,
\begin{align}
I_{22}=-\frac{1}{\pi^3} \sin\frac{\pi C}{2} \sin\pi(A+\frac C2)\sin\pi A \,J(A,B,C).
\end{align}
To calculate $I_{23}$, we do the same trick as for $I_{13}$. The relations
\begin{align}
I_{23}+e^{-i\pi A} \int_1^{\infty} \int_0^1+e^{-i\pi(A+B)} \int_1^{\infty} \int_1^{x_1}+
e^{-i\pi(A+C+B)} \int_1^{\infty} \int_{x_1}^{\infty}=0
\end{align}
and
\begin{align}
I_{23}+e^{-i\pi A} I_{12}
+e^{-i\pi(A+B)} [I_{22}-\int_1^{\infty} \int_{x_1}^{\infty}]+
e^{-i\pi(A+C+B)} \int_1^{\infty} \int_{x_1}^{\infty}=0
\end{align}
hold. In the interval $(x_1,\infty)$, the integral coincides with $I_0(-2A-2B-C-3,B,-A-B-C-2,C,B)$. Hence,
\begin{align}
I_{23}=-\cos\pi A I_{12}-\cos\pi(A+B) I_{22}+[\cos\pi(A+B)-\cos\pi (A+B+C)]\nonumber\\ \cdot
I_0(-2A-2B-C-3,B,-A-B-C-2,C,B).
\end{align}
The integral $I_{33}$ is found from the relation
\begin{align}
\int_{-\infty}^0 \int_{-\infty}^{x_1}+e^{-i\pi C} \int_{-\infty}^0 \int_{x_1}^0
+e^{-i\pi(C+A)} \int_{-\infty}^0 \int_0^{1}+
e^{-i\pi(A+C+B)} \int_{-\infty}^0 \int_{1}^{\infty}=0.
\end{align}
Taking into account that
\begin{align}
\int_{-\infty}^0 \int_{-\infty}^{x_1}=\int_{-\infty}^0 \int_{x_1}^0=\frac{I_{33}}2,
\end{align}
we obtain
\begin{align}
\frac{1+e^{-i\pi C}}2 I_{33} +e^{-i\pi(C+A)} I_{13}+e^{-i\pi(A+C+B)} I_{23}=0
\end{align}
and
\begin{align}
I_{33}=-\frac{1}{\cos\frac{\pi C}2}
\left[\cos\pi (A+\frac{C}{2}) I_{13}+ \cos\pi(A+B+\frac C2) I_{23}\right].
\end{align}
In the calculation of the boundary integrals, we have thus obtained the expressions
\begin{align}
I_{11}&=-\frac{1}{\pi^3} \sin\frac{\pi C}{2} \sin\pi(A+B+\frac C2)\sin\pi(A+B+C)\, J(A,B,C), \nonumber\\
I_{12}&=-\frac{1}{2\pi^3} \sin\pi C \sin\pi(A+B+\frac C2)\sin\pi(A+\frac C2)\, J(A,B,C),  \nonumber\\
I_{13}&=-\frac{1}{2\pi^3} \sin\pi C \sin\pi(A+B+\frac C2)\sin\pi(B+\frac C2)\, J(A,B,C),  \nonumber\\
I_{22}&=-\frac{1}{\pi^3} \sin\frac{\pi C}{2} \sin\pi(A+\frac C2)\sin\pi A \, J(A,B,C), \\
I_{23}&=-\frac{1}{2\pi^3} \sin\pi C \sin\pi(A+\frac C2)\sin\pi(B+\frac C2)\, J(A,B,C),  \nonumber\\
I_{33}&=-\frac{1}{\pi^3} \sin\frac{\pi C}{2} \sin\pi(B+\frac C2)\sin\pi B \, J(A,B,C).\nonumber
\end{align}

\end{document}